%% file: mochejska.tex
\documentclass[10pt,preprint2]{aastex}
\shorttitle{Variables in NGC 2158}
\shortauthors{Mochejska et al.}
\begin{document}

\title{Planets in Stellar Clusters Extensive Search. II.~Discovery of
57 Variables in the Cluster NGC~2158 with Millimagnitude Image
Subtraction Photometry.}

\author{B.~J.~Mochejska\altaffilmark{1}, K.~Z.~Stanek,
D.~D.~Sasselov\altaffilmark{2}, A.~H.~Szentgyorgyi, M.~Westover \&
J.~N.~Winn\altaffilmark{3}}
\affil{Harvard-Smithsonian Center for Astrophysics, 60 Garden St.,
Cambridge, MA~02138}
\email{bmochejs, kstanek, sasselov, saint, mwestover,
jwinn@cfa.harvard.edu}
\altaffiltext{1}{Hubble Fellow.}
\altaffiltext{2}{Alfred P. Sloan Research Fellow.}
\altaffiltext{3}{NSF Astronomy \& Astrophysics Postdoctoral Fellow.}

\begin{abstract}  
We have undertaken a long-term project, Planets in Stellar Clusters
Extensive Search (PISCES), to search for transiting planets in open
clusters. NGC~2158 is one of the targets we have chosen -- an
intermediate age, populous, rather metal poor cluster. In this paper
we present the results of a search for variable stars in the data from
the first season of monitoring at the FLWO 1.2 m telescope. This is
the first variability search ever conducted in this cluster. We
present a catalog of 57 variable stars, most with low amplitude
variability. Among the variables is a cataclysmic variable (CV) which
underwent a 2.5 mag outburst. If it is a member of NGC~2158, this
would be the fourth CV known in an open cluster. We have also found
five $\delta$ Scuti stars, three of which we have two or more
detectable modes of pulsation. Of the 57 variables discovered, 28 have
$R$-band amplitudes of 5\% or below. Six of those vary at or below the
2\% level, including one with 0.08\% variability.
\end{abstract} 
\keywords{ binaries: eclipsing -- cataclysmic variables -- stars: 
variables: $\delta$ Scuti -- color-magnitude diagrams }

\section{{\sc Introduction}}

We have undertaken a long-term project, Planets in Stellar Clusters
Extensive Search (PISCES), to search for transiting planets in open
clusters. In Mochejska et al.\ (2002, hereafter Paper~I) we
demonstrated that we are capable of attaining the requisite
photometric precision, based on one season of data for NGC~6791. 

In this paper we present the results of a search for variable stars in
our second target, NGC~2158, based on the data from the first
observing season. 

NGC~2158 is a rich open cluster of intermediate age (2-3 Gyr). It is
rather metal poor ([Fe/H]=-0.46) and is located at a distance of 3600
pc (Carraro et al.\ 2002, hereafter Ca02; Christian, Heasley and Janes
1985). Despite being one of the most populous galactic open clusters,
it has never been investigated for variability. The only published CCD
photometry for this cluster is that of Christian et al.\ (1985),
Piersimoni et al.\ (1993) and Ca02. An extensive catalog of
photographic photometry and proper motions has been published by
Kharchenko, Andruk \& Schilbach (1997, hereafter KAS97).

The paper is arranged as follows: \S 2 describes the observations, \S
3 summarizes the data reduction procedure and \S 4 contains the
variable star catalog. Concluding remarks are found in \S 5.

\section{{\sc Observations}} 
\begin{figure}[t]
\plotone{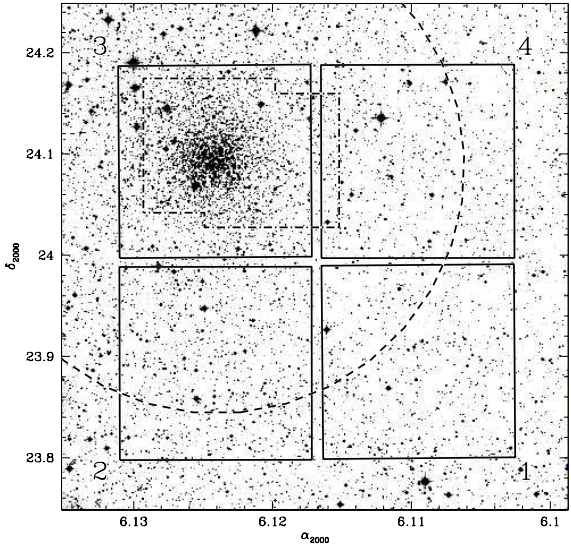}
\caption{Digital Sky Survey image of NGC~2158 showing the field of
view of the 4Shooter. The chips are numbered clockwise from 1 to 4
starting from the bottom left chip. NGC~2158 is centered on Chip 3.
North is up and east is to the left. The fields covered by KAS97 and Ca02
are plotted with dashed and dot-dashed lines, respectively.}
\label{chips}
\end{figure}

The data were obtained at the Fred Lawrence Whipple Observatory (FLWO)
1.2 m telescope using the 4Shooter CCD mosaic with four thinned, back
side illuminated AR coated Loral $2048^2$ CCDs (Szentgyorgyi et
al.~2004).  The camera, with a pixel scale of $0\farcs 33$
pixel$^{-1}$, gives a field of view of $11\farcm 4\times 11\farcm 4$
for each chip. The cluster was centered on Chip 3
(Fig.~\ref{chips}). The data were collected during 20 nights, from 3
January to 11 March, 2003. A total of $250\times 900$~s $R$ and
$69\times 450$~s $V$-band exposures were obtained. The median seeing
was $1\farcs 9$ in $R$ and $2\farcs 1$ in $V$.

\section{{\sc Data Reduction}}

\subsection{{\it Image Subtraction Photometry}}

The preliminary processing of the CCD frames was performed with the
standard routines in the IRAF ccdproc package.\footnote{IRAF is
distributed by the National Optical Astronomy Observatories, which are
operated by the Association of Universities for Research in Astronomy,
Inc., under cooperative agreement with the NSF.}

Photometry was extracted using the ISIS image subtraction package
(Alard \& Lupton 1998, Alard 2000). A brief outline of the applied
reduction procedure is presented here. For a more detailed description
the reader is referred to Paper~I.

The ISIS reduction procedure consists of the following steps: (1)
transformation of all frames to a common $(x,y)$ coordinate grid; (2)
construction of a reference image from several of the best exposures;
(3) subtraction of each frame from the reference image; (4) selection
of stars to be photometered and (5) extraction of profile photometry
from the subtracted images.

All computations were performed with the frames internally subdivided
into four sections ({\tt sub\_x=sub\_y=2}). Differential brightness
variations of the background were fit with a first degree polynomial
({\tt deg\_bg=1}). A convolution kernel varying quadratically with
position was used ({\tt deg\_spatial=2}). The psf width ({\tt
psf\_width}) was set to 33 pixels and the photometric radius ({\tt
radphot}) to 5 pixels. The reference images were constructed from
20 best exposures.

\begin{figure*}[ht]
\epsscale{1.5}
\plotone{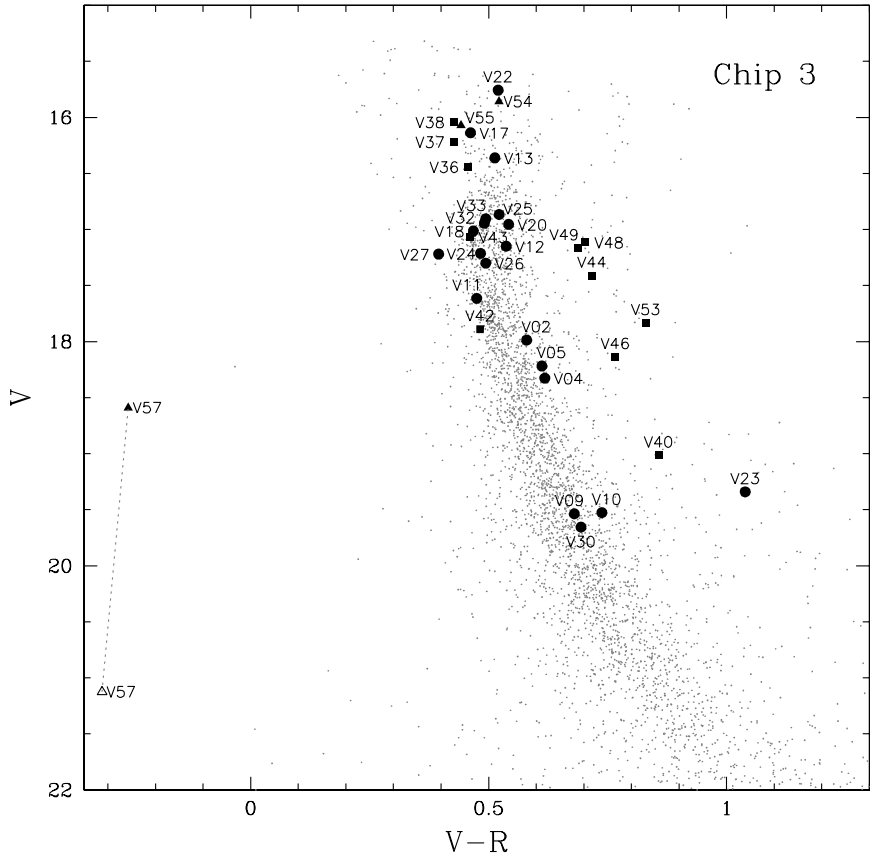}
\caption{$V/\vr$ color-magnitude diagram (CMD) for Chip 3, centered on
NGC~2158. Eclipsing binaries are plotted with circles, other periodic
variables with squares and miscellaneous variables with triangles.
The CV candidate, V57, is plotted at maximum and minimum (filled and
open symbol, respectively). There are 3842 stars above $R=21$.}
\label{cmd3}
\end{figure*}

\begin{figure*}[ht]
\epsscale{2}
\plotone{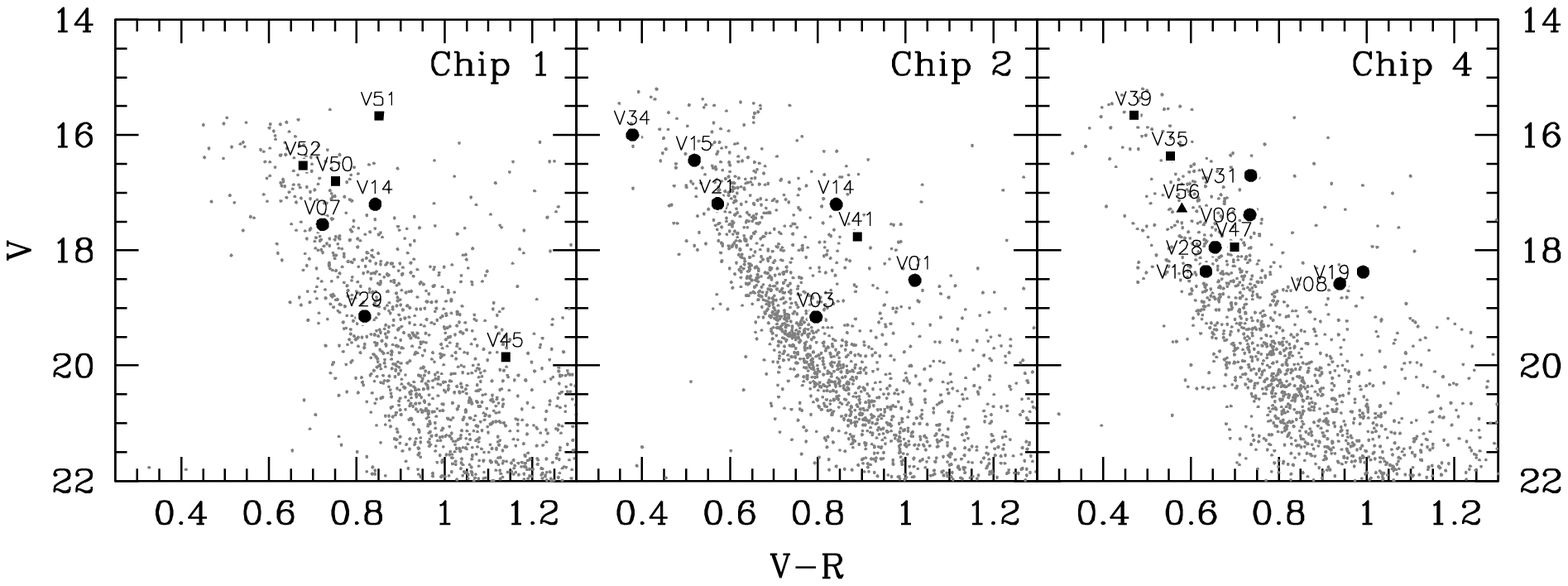}
\caption{$V/\vr$ CMDs for Chips 1, 2 and 4. Eclipsing binaries are
plotted with circles, other periodic variables with squares and
miscellaneous variables with triangles. There are 1429, 1764 and 1436
stars above $R=21$ in Chips 1, 2 and 4, respectively.}
\label{cmd124}
\end{figure*}

\subsection{{\it Calibration}}
The $VR$ photometry in Chips 3 and 4 was calibrated from the
photometry published by Ca02. 1410 stars above $R=18.5$ and 1312 above
$V=19$ were used to determine the zero point of the magnitude scale
for Chip 3. Since the overlap with Chip 4 was marginal (see
Fig.~\ref{chips}), we used 44 stars above $R=19$ and $V=20$. The rms
scatter of the residuals in $R$ and $V$, respectively, was 0.046 and
0.052 in Chip 3 and 0.065 and 0.059 in Chip 4.

The $R$-band photometry for Chips 1 and 2 was calibrated from the KAS97
photographic photometry. To determine the zero point we used stars
above 17 mag in both bands: 39 and 19 in $R$ and $V$ for Chip 1, and
169 and 87 for Chip 2. The rms scatter of the residuals in $R$ and
$V$, respectively, was 0.18 and 0.14 in Chip 1 and 0.17 and 0.12 in
Chip 2.

As a consistency check we have calibrated Chip 3 and 4 photometry from
the KAS97 catalog and found an offset between the calibration based on
Ca02. We have also compared the KAS97 and Ca02 catalogs directly, and
confirm the difference between their zero points. The offsets range
from 0.11 to 0.19 in $R$ and 0.04 and 0.1 in $V$, depending on the
sample of stars used. Since we could not constrain well the offsets
between KAS97 and Ca02, we have left Chips 3 and 4 on the Ce02 scale, and
Chips 1 and 2 on the KAS97 scale.

In Figures \ref{cmd3} and \ref{cmd124} we present the $V/\vr$
color-magnitude diagrams (CMD) for Chip 3 and Chips 1, 2, 4,
respectively. The CMD for Chip 3, centered on the cluster, shows a
well defined main sequence (MS) from the turnoff down to $V\sim 21$.
The CMDs for the remaining chips are dominated by disk stars.  There
are 1429, 1764, 3842 and 1436 stars above $R=21$ in Chips 1 through 4,
respectively.

\subsection{{\it Astrometry}}
Equatorial coordinates were determined for the $R$-band template star
lists. The transformation from rectangular to equatorial coordinates
was derived using 740, 922, 1170, and 777 transformation stars from
the USNO B-1 catalog (Monet et al.\ 2003) in Chips 1 through 4,
respectively. The average difference between the catalog and the
computed coordinates for the transformation stars was $0\farcs 09-
0\farcs 11$.

\subsection{{\it Selection of variables}}

We extracted the light curves for all stars detected by DAOphot
(Stetson 1987) on the template frames and searched them for
variability using the index $J$ (Stetson 1996), as described in more
detail in Paper~I.  The definition of $J$ requires observations to be
closely spaced in time. Points separated by a value smaller than some
threshold (0.03 days in our case) are grouped into pairs and only when
both have a residual from the mean of the same sign, the pair
contributes positively to the variability index.

To search for periodicities we have used the method introduced by
Schwarzenberg-Czerny (1996), employing periodic orthogonal polynomials
to fit the observations and the analysis of variance (ANOVA) statistic
to evaluate the quality of the fit.

\begin{figure*}[hp]
\epsscale{2}
\plotone{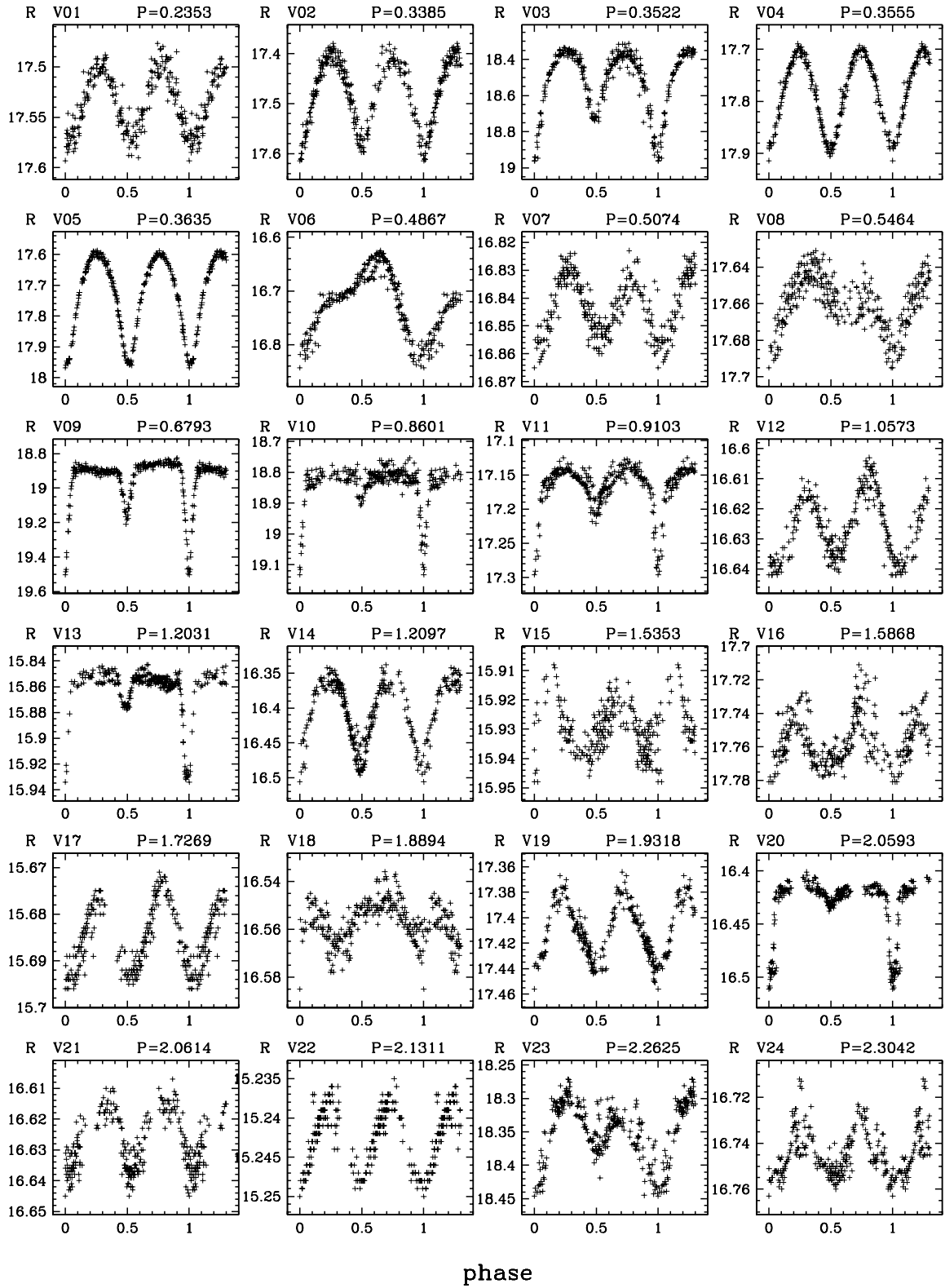}
\caption{The light curves of the 34 eclipsing binaries.}
\label{lc:ecl}
\end{figure*}

\addtocounter{figure}{-1}
\begin{figure*}[ht]
\plotone{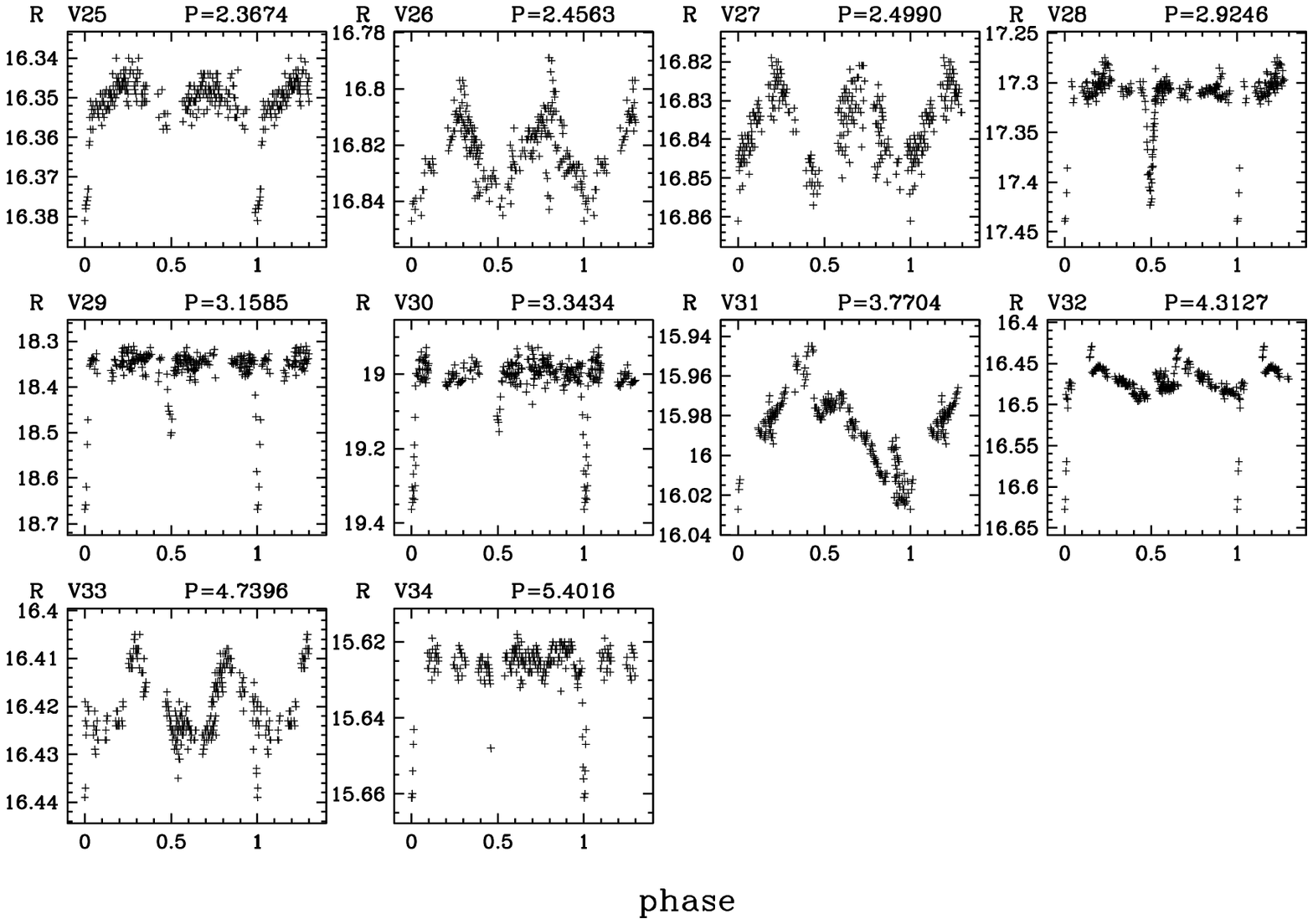}
\caption{Continued.}
\end{figure*}

\section{{\sc Variable Star Catalog}}

We have discovered 57 variables in all four chips, all newly discovered: 34
eclipsing binaries, 19 other periodic and 4 long period or
non-periodic variables. 

\begin{figure}[hp]
\epsscale{1}
\plotone{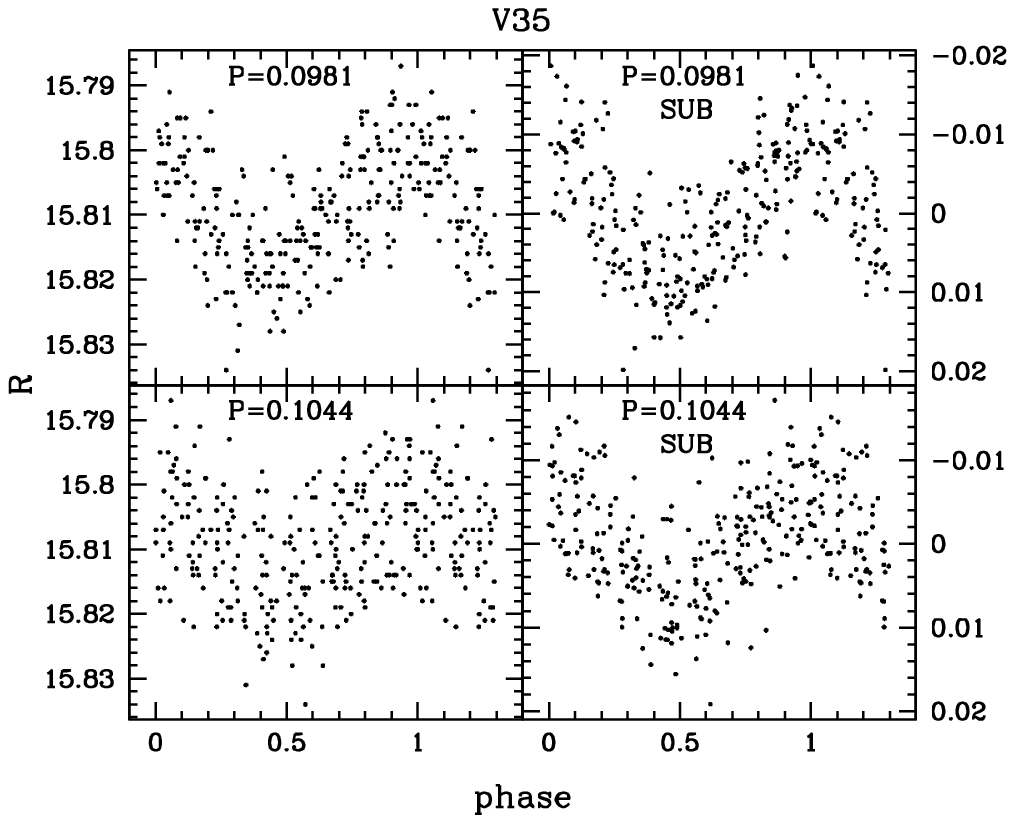}
\caption{The $R$-band light curves of $\delta$ Scuti variable V35,
phased with each of the two detected periods: raw in the left panels,
and subtracted of variability corresponding to other period in the
right panels.}
\label{lc:mm1}
\end{figure}

\begin{figure*}[ht]
\epsscale{2}
\plotone{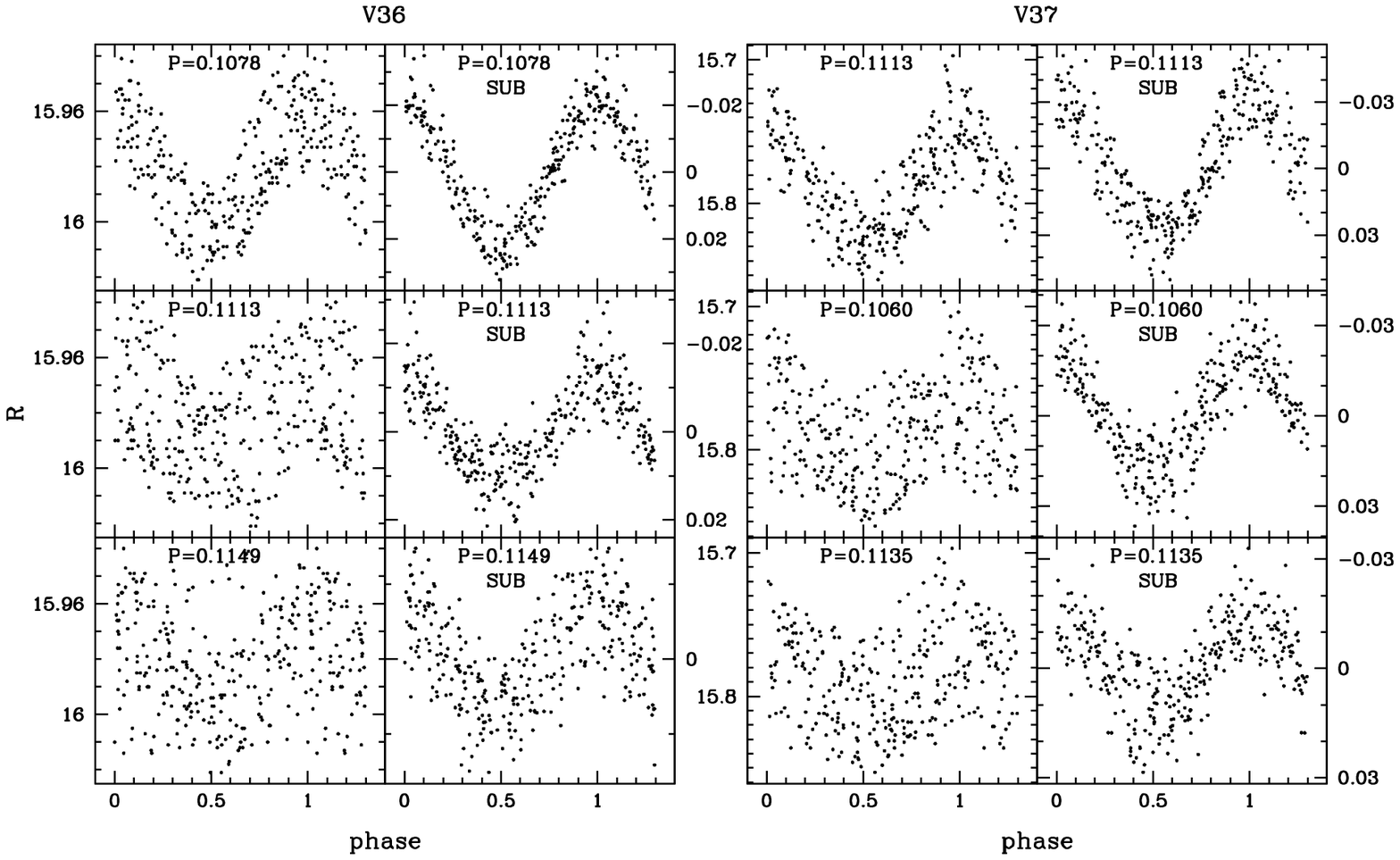}
\caption{The $R$-band light curves of $\delta$ Scuti variables V36 and
V37, phased with each of the three detected periods: raw in the left
panels, and subtracted of variability corresponding to other periods
in the right panels.}
\label{lc:mm2}
\end{figure*}

The properties of the variables are discussed in \S 4.1-4.5. Tables
\ref{tab:ecl}, \ref{tab:pul} and \ref{tab:misc} list the following
parameters for the variables: the identification number, right
ascension, declination, period (if applicable), $R$ and $V$-band
magnitudes (maximum for non-periodic and eclipsing variables, flux
weighted mean for periodic) and variability amplitudes in each band
(semiamplitudes for periodic variables). The last two columns in
Tab.~\ref{tab:ecl} and \ref{tab:pul} provide the probabilities $P_1$
and $P_2$ for cluster core and corona membership, taken from KAS97. The
$R$-band light curves of all variables are shown in Figures
\ref{lc:ecl}-\ref{lc:misc}.\footnote{The $VR$ band
photometry and finding charts for all variables are available from the
authors via the anonymous ftp on cfa-ftp.harvard.edu, in the
/pub/bmochejs/PISCES directory.}

The variables are plotted on CMDs for their respective chips (Figs
\ref{cmd3} and \ref{cmd124}). Eclipsing binaries are marked with
circles, other periodic variables with squares and miscellaneous
variables with triangles. By far the largest number of variables, 34,
were found on Chip 3, centered on the cluster, compared to 6, 7 and 10
on Chips 1, 2 and 4, respectively. This is in agreement with the
5$\arcmin$ radius estimate for the cluster (Lyng\r{a} 1987).

\subsection{Eclipsing Binaries} 
We have identified 34 eclipsing binaries. Six are contact W UMa type
systems (V01, V02, V04, V05, V07, V12). More than half of the
variables could either be EB-type eclipsing binaries or ellipsoidal
variables, as most have periods in excess of a day and amplitudes
below 0.1 mag (V03, V08, V11, V14-V27, V33). It is possible that some
of the longer period systems could be BY Dra type spotted stars with
half the reported period. There are also seven Algol type detached
binaries (V09, V10, V13, V28-V30, V32, V34). V06 and probably V31 are
RS CVn type binaries, where one of the components has non-uniform
surface brightness due to cool spots. A starspot wave is seen on top
of the variability due to eclipses.

Proper motion data from KAS97 show that V12, V22 and V24 are most likely
cluster members, while V14 is rather not. For V21, V28 and V31 the
data are inconclusive.

\subsection{$\delta$ Scuti stars}

We have identified five variables, V35-V39, with periods 0.0981-0.1220
d and amplitudes 0.008-0.076 mag, located in the blue straggler region
of the CMD. Their periods and amplitudes are typical for $\delta$
Scuti variables. Their positions on the CMD are also consistent with
their identification as $\delta$ Scuti variables. In clusters
considerably older than 1 Gyr the main sequence turnoff is located
redward of the instability strip. Consequently, only blue stragglers
can display $\delta$ Scuti variability (Rodr{\'{\i}}guez \& Breger
2001).

For three of these variables, V35-V37, we have detected pulsations in
more than one mode. Table~\ref{tab:mm} lists identified
periods. Figures~\ref{lc:mm1} and \ref{lc:mm2} show their light
curves, phased with each detected period: raw in the left panels, and
subtracted of variability corresponding to other periods in the right
panels. The simultaneous excitation of multiple modes of pulsation is
typical for $\delta$ Scuti variables, and such stars have been
detected in open clusters (Rodr{\'{\i}}guez 2002, Freyhammer, Arentoft
\& Sterken 2001, Pigulski, Kolaczkowski, \& Kopacki 2000).

The ratios of the periods range from 1.020 to 1.071. Such close
frequency pairs are observed in the majority of well studied $\delta$
Scuti stars (Breger \& Bischof 2002). For periods around 0.1 d the
ratio between the fundamental and first overtone period for Population
I $\delta$ Scuti stars is generally below 0.775 (Templeton, Basu \&
Demarque 2002). This implies that at least one of the excited modes of
pulsation is non-radial.

Using the fundamental mode Period-Luminosity (P-L) relation for
$\delta$ Scuti stars (Eq.~4 in Petersen \& Christensen-Dalsgaard
1999), $R_V=3.2$ (Patriarchi, Morbidelli, \& Perinotto 2003),
$E(\bv)=0.60$\footnote{The final determination of $E(\bv)$ in Ca02 was
$0.55\pm0.10$ mag. Here we are using the reddening and distance
modulus obtained by them from isochrone fits.} and $(m-M)_0=12.80$
(Ca02), we get deviations from the cluster distance modulus less than
0.15 mag for all stars except V39 (Tab.~\ref{tab:mm}). This confirms
that these stars are most likely $\delta$ Scutis belonging to
NGC~2158. One possible cause of discrepancy for V39 could be that the
mode of pulsations we have identified is not the fundamental mode.

Four of these stars have proper motion data in KAS97. Cluster membership
of V36 is confirmed with high probability, while for V35 and V39 the
data are not conclusive. The catalog reports a very low cluster
membership probability for V38, which is in disagreement with our
conclusions based on the $\delta$ Scuti P-L relation. The study of KAS97
was also targeting much brighter stars ($V>8$ mag) in M~35 and two
other clusters at smaller distances than NGC~2158. The authors say
that their proper motion measurement ``accuracy decreases rapidly for
stars fainter than $V=15.5$ mag'', while the $V$ magnitude range of
NGC~2158 variables is 15.7-18.0, and $V=16.04$ for V38. 

\begin{figure*}[ph]
\plotone{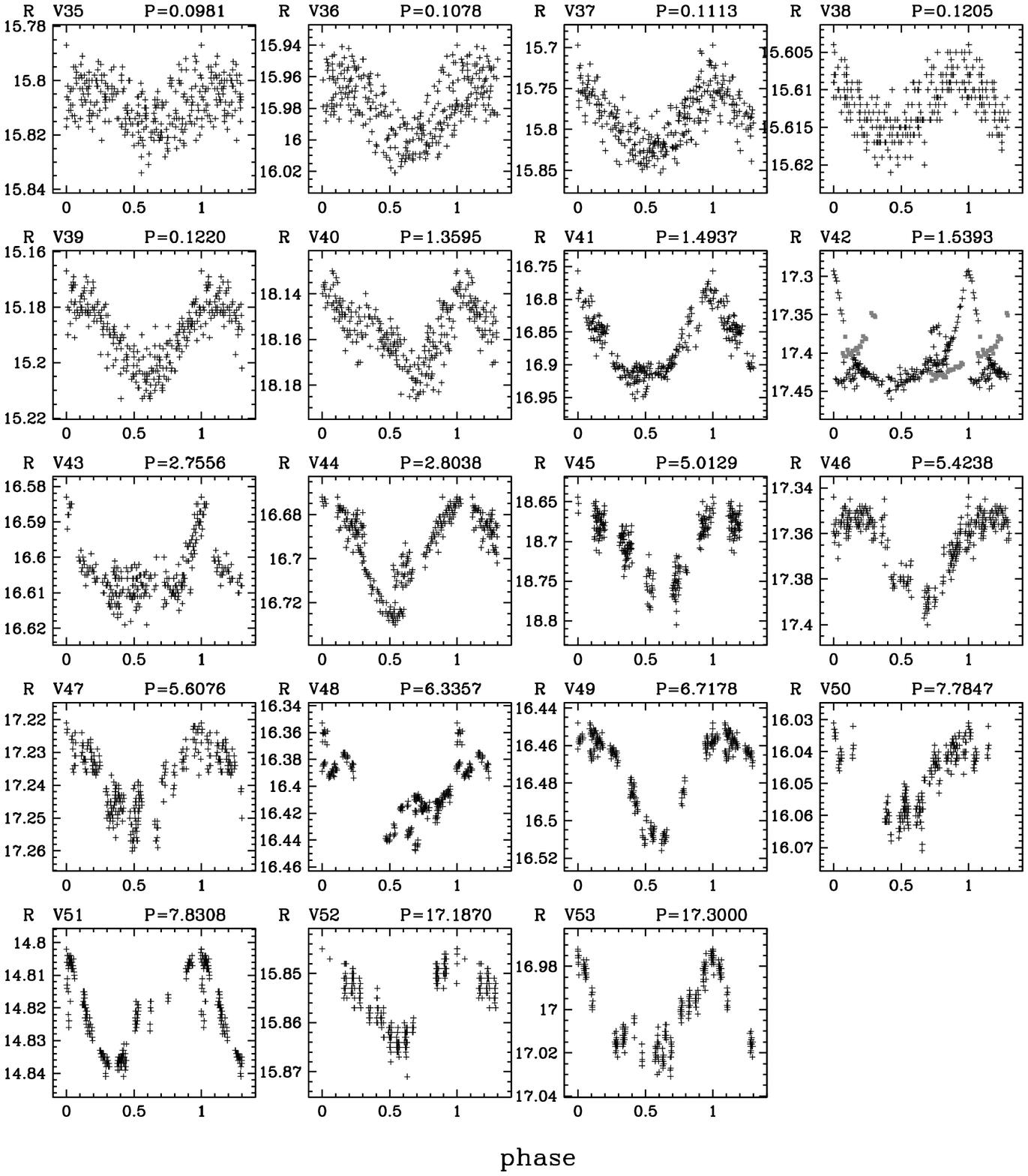}
\caption{The light curves of the 19 other periodic variables.}
\label{lc:pul}
\end{figure*}

\subsection{Other periodic variables}

All of the remaining periodic variables are located redward of the MS,
except for V42 and V43, which are located at its blue edge. Some of
these variables could be BY Draconis type variables -- rotating
spotted stars. Six of the 13 variables are located on Chips 1, 2 and
4, making it likely that many of them are not cluster members.

V48 seems to have a variable amplitude. Variables V44, V48 and V49 on
Chip 3, if they belong to the cluster, might be members of the newly
proposed class of variable stars termed ``red stragglers'' (Albrow et
al.\ 2001) or ``sub-subgiant stars'' (Mathieu et al.\ 2003).  KAS97
gives a probability of corona memberships of 80\%, 69\% and 41\% for
V44, V48 and V49, respectively. To date, six such stars have been
found in 47 Tuc (Albrow et al.\ 2001) and two in M67 (Mathieu et al.\
2003). Thus far, the origin and evolutionary status of these stars
remains unknown.

Variables V50 and V52 on Chip 1 and the RS CVn eclipsing binary V31 on
Chip 4 also fall in the same region in the CMD. Since they are further
from the cluster center, it is less likely they are members of
NGC~2158 and of this class. The KAS97 catalog suggests V50 is most likely
not a cluster member, and it is inconclusive for V31.

The variable V42 has a sharply peaked light curve, reminiscent of RR
Lyrae or short-period Cepheids. Its amplitude of 0.14 in $R$ is much
too low for either of those two types of variables, unless it is
blended with another star along the same line of sight. The KAS97
catalog gives it a high probability of being a cluster member. If V42
were really a member, it would have $M_V\sim3.2$ and would be too
faint to be either of these variables. On the CMD it is located at the
blue edge of the MS (Fig.~\ref{cmd3}). The data for V42 from all
nights, except the sixth and seventh, phase well with a period of
1.5393 days. In Fig.~\ref{lc:pul} the points from these two discrepant
nights are marked with solid gray squares.  For the full dataset (or
even nights 1-7) a good period cannot be determined. The examination
of images from nights six and seven did not reveal any defects, such
as bad columns, extremely high background or a bad psf, or anything
that would suggest that the brightness measurements might be less
accurate. Also, the data from these two nights show a coherent rise in
brightness, which makes it more unlikely that these measurements are
erroneous.

\begin{figure*}[ht]
\plotone{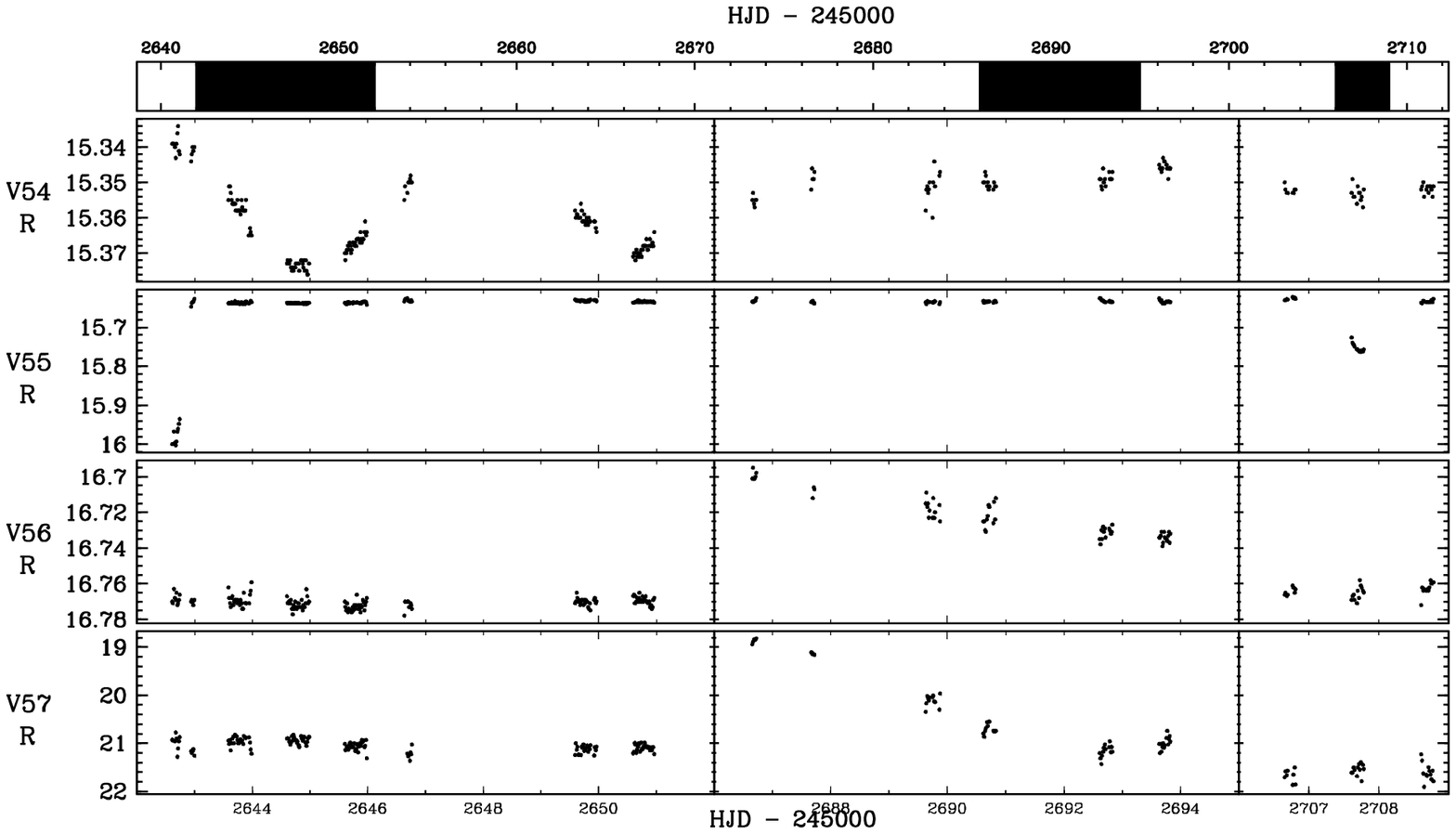}
\caption{The light curves of the 4 miscellaneous variables. The top
window illustrates the distribution in time of the three subwindows
plotted for the variables.}
\label{lc:misc}
\end{figure*}

\subsection{Cataclysmic Variable Candidate}

We have identified a cataclysmic variable (CV) candidate V57. The star
has undergone an outburst with an amplitude of 2.5 mag in $R$ and
$V$. Its very blue color, $\vr=-0.26$ in maximum and $-0.31$ in
minimum, supports this classification. If confirmed spectroscopically
as a CV, and as a member of NGC~2158, this would be only the fourth CV
known in an open cluster, with the other three being star 186 in M67
(Gilliland et al.\ 1991) and B7 and B8 in NGC~6791 (Kaluzny et al.\
1997).

The $\vr$ color of this CV candidate is in fact implausibly blue.
We searched for this star in the Ca02 catalog, to check its color, but
it is not listed there. We have obtained independent color derivations
for this variable from three sets of $R$ and $V$-band images taken on
different nights when this object was near maximum brightness. One of
these sets came from our most recent observations, not included in
this analysis. These sets were calibrated independently against the
Ca02 catalog. The $\vr$ color estimates ranged from -0.24 to
-0.31. Unfortunately the seeing in $V$ was over $2\arcsec$ on all
$V$-band images when the variable was at maximum, so the $V$-band
magnitude probably has a considerable error associated with it.

Assuming the NGC~2158 distance and reddening derived by Ca02, we get
the minimum absolute $V$-band magnitude $M_V(min)$ of 6.4 mag. From
Fig.~3.5 in Warner (1995) it is apparent that $M_V(min)$ is roughly 8
mag for U Gem type dwarf novae, but even the most extreme Z Cam type
systems have $M_V(min)$ below 6.8 mag. This would imply that V57 is a
dwarf nova located in front of the cluster, and would also make the
color less extreme, assuming it has a much smaller reddening.

It should be noted, however, that the $V$ band magnitude alone is not
sufficient to exclude this variable as a cluster member. Another
cataclysmic variable, B7 in NGC~6791, is seen most of the time in a
high state at $M_V\sim5$, and has been caught only once in its low
state at $M_V(min)=7.4$. Further investigation is needed to resolve
the nature of V57. A spectrum of a dwarf nova in quiescence should
display Balmer emission on a blue continuum. During outburst the
emission lines are gradually overwhelmed by the increasing continuum
and development of broad absorption lines (Warner 1995). If V57 is in
a ``high'' level, like B7, then this will be apparent in the spectrum
and emission lines should not be prominent.

\subsection{Long period or non-periodic variables}

The variable V54 exhibits irregular variability and is located in the
blue straggler region of the CMD. V55 is a detached eclipsing binary
with one primary and one secondary eclipse observed, also probably a
blue straggler. 

The shape of the light curve of V56 is very similar to V57, only with
a much smaller amplitude of 0.07 in $R$ and $0.05$ in $V$. On the CMD
it is located near the main sequence turnoff. We have considered the
possibility that the variability of V56 could be an artifact related
to V57, although the fact that V56 was detected on a different chip than
V57 makes it less likely. We have examined images from two nights,
when the variable was at minimum and maximum brightness, and found
nothing suspicious. A simple examination under IRAF with imexam
confirms the variability of this star, as well as its amplitude.

\section{{\sc Conclusions}}

In this paper we present $VR$ photometry for 57 newly discovered
variable stars in the intermediate age, populous, rather metal poor
cluster NGC~2158. We have identified 34 eclipsing binaries, 19 other
periodic variables and 4 long period or non-periodic variables. Among
these variables we have discovered five $\delta$ Scuti stars. For
three of them we have identified two or more modes of pulsation.  We
have also found a cataclysmic variable which underwent a 2.5 mag
outburst.  If it is a member of NGC~2158 (which at this point does not
seem very likely) this would be the fourth CV known in an open
cluster.

Of the 57 variables discovered, 28 have $R$-band amplitudes of 5\% or
below. Six of those vary at or below the 2\% level, including one with
0.08\% variability. This, together with the results from Paper~I,
shows that we are capable of attaining requisite photometric precision
to detect variability at the level expected for planetary transits.

\acknowledgments{ This research has made use of the USNOFS Image and
Catalogue Archive operated by the United States Naval Observatory,
Flagstaff Station (http://www.nofs.navy.mil/data/fchpix/), the Digital
Sky Survey, produced at the Space Telescope Science Institute under
U.S. Government grant NAG W-2166 and the WEBDA open cluster database
maintained by J.~C.~Mermilliod (http://obswww.unige.ch/webda/).
Support for BJM was provided by NASA through Hubble Fellowship grant
HST-HF-01155.01-A from the Space Telescope Science Institute, which is
operated by the Association of Universities for Research in Astronomy,
Incorporated, under NASA contract NAS5-26555. JNW gratefuly
acknowledges the support of NSF grant AST-0104347.  KZS acknowledges
support from the William F. Milton Fund.}

\input{mochejska.tab1.tex}
\input{mochejska.tab2.tex}
\input{mochejska.tab3.tex}
\input{mochejska.tab4.tex}

\end{document}

%% file: mochejska.tab1.tex
\begin{deluxetable}{rrrrrrrrlccccc}
\tabletypesize{\footnotesize}
\tablewidth{0pc}
\tablecaption{Eclipsing binaries in NGC 2158}
\tablehead{\colhead{ID} & \colhead{$\alpha_{2000}$ [h]} &
\colhead{$\delta_{2000}$ [$\circ$]} &\colhead{P [d]} & \colhead{$R_{max}$} &
\colhead{$V_{max}$} &\colhead{$A_R$} & \colhead{$A_V$}&
\colhead{$P_1$} &\colhead{$P_2$}}
\startdata
 V01 &  6 07 32.6 & 23 49 21.2 &  0.2353 & 17.498 & 18.519  &  0.092 &  0.083 & \nodata & \nodata \\
 V02 &  6 07 23.3 & 24 06 12.4 &  0.3385 & 17.406 & 17.986  &  0.206 &  0.202 & \nodata & \nodata \\
 V03 &  6 07 04.9 & 23 48 53.4 &  0.3522 & 18.360 & 19.156  &  0.595 &  0.714 & \nodata & \nodata \\
 V04 &  6 07 17.5 & 24 04 45.7 &  0.3555 & 17.709 & 18.327  &  0.198 &  0.211 & \nodata & \nodata \\
 V05 &  6 07 40.6 & 24 05 03.7 &  0.3635 & 17.606 & 18.218  &  0.356 &  0.396 & \nodata & \nodata \\
 V06 &  6 06 31.3 & 24 00 22.1 &  0.4867 & 16.649 & 17.383  &  0.191 &  0.281 & \nodata & \nodata \\
 V07 &  6 06 37.0 & 23 50 41.2 &  0.5074 & 16.830 & 17.552  &  0.033 &  0.040 & \nodata & \nodata \\
 V08 &  6 06 44.8 & 24 06 57.9 &  0.5464 & 17.641 & 18.580  &  0.053 &  0.079 & \nodata & \nodata \\
 V09 &  6 07 29.4 & 24 10 06.4 &  0.6793 & 18.857 & 19.537  &  0.638 &  0.691 & \nodata & \nodata \\
 V10 &  6 07 34.9 & 24 04 25.7 &  0.8601 & 18.788 & 19.526  &  0.331 &  0.285 & \nodata & \nodata \\
 V11 &  6 07 21.4 & 24 05 39.8 &  0.9103 & 17.140 & 17.615  &  0.151 &  0.136 & \nodata & \nodata \\
 V12 &  6 07 18.7 & 24 06 50.1 &  1.0573 & 16.612 & 17.149  &  0.030 &  0.036 & 71 & 78  \\
 V13 &  6 07 28.1 & 24 06 35.3 &  1.2031 & 15.849 & 16.362  &  0.083 &  0.090 & \nodata & \nodata \\
 V14 &  6 07 27.2 & 23 52 24.5 &  1.2097 & 16.360 & 17.202  &  0.139 &  0.125 &  1 & 28  \\
 V15 &  6 07 36.1 & 23 48 24.1 &  1.5353 & 15.920 & 16.439  &  0.027 &  0.025 & \nodata & \nodata \\
 V16 &  6 06 36.6 & 24 03 29.2 &  1.5868 & 17.735 & 18.369  &  0.046 &  0.059 & \nodata & \nodata \\
 V17 &  6 07 31.9 & 24 06 00.3 &  1.7269 & 15.676 & 16.138  &  0.020 &  0.029 & \nodata & \nodata \\
 V18 &  6 07 31.0 & 24 05 50.5 &  1.8894 & 16.546 & 17.014  &  0.034 &  0.039 & \nodata & \nodata \\
 V19 &  6 06 20.5 & 24 04 48.0 &  1.9318 & 17.385 & 18.377  &  0.067 &  0.080 & \nodata & \nodata \\
 V20 &  6 07 20.6 & 24 06 01.3 &  2.0593 & 16.413 & 16.955  &  0.097 &  0.102 & \nodata & \nodata \\
 V21 &  6 07 36.3 & 23 56 21.4 &  2.0614 & 16.616 & 17.188  &  0.029 &  0.033 & 10 & 47  \\
 V22 &  6 07 37.7 & 24 07 40.2 &  2.1311 & 15.238 & 15.758  &  0.011 &  0.013 & 72 & 90  \\
 V23 &  6 07 03.5 & 24 01 45.3 &  2.2625 & 18.301 & 19.340  &  0.144 &  0.210 & \nodata & \nodata \\
 V24 &  6 07 43.4 & 24 06 22.9 &  2.3042 & 16.731 & 17.214  &  0.030 &  0.040 & 59 & 79  \\
 V25 &  6 07 19.9 & 24 06 24.4 &  2.3674 & 16.345 & 16.867  &  0.034 &  0.031 & \nodata & \nodata \\
 V26 &  6 07 24.1 & 24 07 53.9 &  2.4563 & 16.807 & 17.301  &  0.039 &  0.065 & \nodata & \nodata \\
 V27 &  6 07 30.2 & 24 07 50.6 &  2.4990 & 16.825 & 17.220  &  0.032 &  0.071 & \nodata & \nodata \\
 V28 &  6 06 50.0 & 24 08 26.2 &  2.9246 & 17.295 & 17.950  &  0.138 &  0.130 & 15 & 59  \\
 V29 &  6 06 46.2 & 23 49 41.4 &  3.1585 & 18.329 & 19.147  &  0.320 &  0.388 & \nodata & \nodata \\
 V30 &  6 07 14.9 & 24 09 40.9 &  3.3434 & 18.961 & 19.655  &  0.387 &  0.455 & \nodata & \nodata \\
 V31 &  6 06 49.0 & 24 01 43.9 &  3.7704 & 15.967 & 16.703  &  0.059 &  0.057 & 11 & 47  \\
 V32 &  6 07 27.9 & 24 06 22.8 &  4.3127 & 16.454 & 16.945  &  0.154 &  0.154 & \nodata & \nodata \\
 V33 &  6 07 33.3 & 24 04 34.5 &  4.7396 & 16.411 & 16.905  &  0.026 &  0.040 & \nodata & \nodata \\
 V34 &  6 07 19.9 & 23 49 51.6 &  5.4016 & 15.621 & 15.999  &  0.040 &  0.033 & \nodata & \nodata \\
\enddata
\tablecomments{Cluster core and corona membership probabilities $P_1$ and $P_2$ taken from K97.}
\label{tab:ecl}
\end{deluxetable}

%% file: mochejska.tab2.tex
\begin{deluxetable}{rrrrrrrrlccccc}
\tabletypesize{\footnotesize}
\tablewidth{0pc}
\tablecaption{Other periodic variables in NGC 2158}
\tablehead{\colhead{ID} & \colhead{$\alpha_{2000}$ [h]} &
\colhead{$\delta_{2000}$ [$\circ$]} &\colhead{P [d]} & \colhead{$\langle R\rangle$} &
\colhead{$\langle V\rangle$} &\colhead{$A_R$} & \colhead{$A_V$}&
\colhead{$P_1$} &\colhead{$P_2$}}
\startdata
 V35 &  6 06 47.1 & 24 07 41.6 &  0.0981 & 15.809 & 16.362  &  0.008 &  0.012 & 13 & 58  \\
 V36 &  6 07 16.8 & 24 05 44.9 &  0.1078 & 15.983 & 16.440  &  0.021 &  0.031 & 83 & 90  \\
 V37 &  6 07 26.0 & 24 04 23.7 &  0.1113 & 15.790 & 16.217  &  0.038 &  0.045 & \nodata & \nodata \\
 V38 &  6 07 11.3 & 24 03 36.5 &  0.1205 & 15.612 & 16.039  &  0.004 &  0.006 &  3 &  4  \\
 V39 &  6 06 50.6 & 24 10 56.6 &  0.1220 & 15.191 & 15.661  &  0.013 &  0.019 &  4 & 23  \\
 V40 &  6 07 05.6 & 24 07 06.5 &  1.3595 & 18.158 & 19.015  &  0.017 &  0.016 & \nodata & \nodata \\
 V41 &  6 07 04.2 & 23 58 09.6 &  1.4937 & 16.869 & 17.759  &  0.061 &  0.045 &  7 & 28  \\
 V42 &  6 07 36.3 & 24 02 07.5 &  1.5393 & 17.407 & 17.889  &  0.070 &  0.087 & 67 & 88  \\
 V43 &  6 07 49.9 & 24 09 45.6 &  2.7556 & 16.605 & 17.065  &  0.010 &  0.011 & \nodata & \nodata \\
 V44 &  6 07 05.8 & 24 08 51.4 &  2.8038 & 16.695 & 17.412  &  0.023 &  0.026 & 44 & 80  \\
 V45 &  6 06 27.8 & 23 49 58.7 &  5.0129 & 18.712 & 19.851  &  0.050 &  0.069 & \nodata & \nodata \\
 V46 &  6 07 09.7 & 24 06 50.3 &  5.4238 & 17.370 & 18.136  &  0.019 &  0.031 & \nodata & \nodata \\
 V47 &  6 06 40.4 & 24 06 34.4 &  5.6076 & 17.239 & 17.938  &  0.012 &  0.011 &  7 & 51  \\
 V48 &  6 07 06.2 & 24 02 10.3 &  6.3357 & 16.410 & 17.113  &  0.029 &  0.041 & 41 & 69  \\
 V49 &  6 07 10.2 & 24 10 19.5 &  6.7178 & 16.478 & 17.167  &  0.027 &  0.029 & 26 & 51  \\
 V50 &  6 06 43.0 & 23 55 15.3 &  7.7847 & 16.049 & 16.800  &  0.011 &  0.016 &  0 & 24  \\
 V51 &  6 06 25.1 & 23 55 11.3 &  7.8308 & 14.821 & 15.672  &  0.017 &  0.017 & \nodata & \nodata \\
 V52 &  6 06 24.0 & 23 54 25.7 & 17.1870 & 15.855 & 16.533  &  0.008 &  0.017 & \nodata & \nodata \\
 V53 &  6 07 49.0 & 24 02 02.8 & 17.3000 & 17.004 & 17.834  &  0.020 &  0.025 & 34 & 65  \\
\enddata
\tablecomments{Cluster core and corona membership probabilities $P_1$ and $P_2$ taken from K97.}
\label{tab:pul}
\end{deluxetable}

%% file: mochejska.tab3.tex
\begin{deluxetable}{rrrrrrrlccc}
\tabletypesize{\footnotesize}
\tablewidth{0pc}
\tablecaption{Miscellaneous variables in NGC 2158}
\tablehead{\colhead{ID} & \colhead{$\alpha_{2000}$ [h]} &
\colhead{$\delta_{2000}$ [$\circ$]} & \colhead{$R_{max}$} &
\colhead{$V_{max}$} &\colhead{$A_R$} &\colhead{$A_V$}}
\startdata
 V54 &  6 07 07.0 & 24 05 25.4 & 15.337 & 15.859  &  0.034 &  0.033  \\
 V55 &  6 07 25.7 & 24 05 45.7 & 15.630 & 16.072  &  0.370 &  0.340  \\
 V56 &  6 06 17.2 & 24 03 39.8 & 16.699 & 17.278  &  0.074 &  0.052  \\
 V57 &  6 07 33.8 & 24 07 55.2 & 18.847 & 18.590  &  2.590 &  2.535  \\
\enddata
\label{tab:misc}
\end{deluxetable}

%% file: mochejska.tab4.tex
\begin{deluxetable}{ccccc}
\tabletypesize{\footnotesize}
\tablewidth{0pc}
\tablecaption{Properties of $\delta$ Scuti stars in NGC 2158}
\tablehead{\colhead{ID} & \colhead{P$_1$ [d]}& \colhead{P$_2$ [d]} 
&\colhead{P$_3$ [d]} &\colhead{(m-M)$_0$}}
\startdata
V35 & 0.098051 & 0.104419 & \nodata  & 12.65\\ 
V36 & 0.107830 & 0.111341 & 0.114875 & 12.89\\ 
V37 & 0.111285 & 0.106032 & 0.113531 & 12.71\\
V38 & 0.120462 & \nodata  & \nodata  & 12.66\\
V39 & 0.121997 & \nodata  & \nodata  & 12.31\\
\enddata
\label{tab:mm}
\end{deluxetable}

%% file: mochejska.bbl
\begin{references}
\reference{} Alard, C., Lupton, R.\ 1998, ApJ, 503, 325
\reference{} Alard, C.\ 2000, A\&AS, 144, 363
\reference{} Albrow, M.~D., Gilliland, R.~L., Brown, T.~M., Edmonds,  
             P.~D., Guhathakurta, P., \& Sarajedini, A.\ 2001, \apj,
             559, 1060
\reference{} Breger, M.~\& Bischof, K.~M.\ 2002, \aap, 385, 537
\reference{} Carraro, G., Girardi, L., \& Marigo, P.\ 2002, \mnras,
             332, 705 (Ca02)
\reference{} Christian, C.~A., Heasley, J.~N., \& Janes, K.~A.\ 1985,
             \apj, 299, 683
\reference{} Freyhammer, L.~M., Arentoft, T., \& Sterken, C.\ 2001,
             \aap, 368, 580
\reference{} Gilliland, R.~L.~et al.\ 1991, \aj, 101, 541 
\reference{} Kaluzny, J., Stanek, K.~Z., Garnavich, P.~M. \& Challis, P.\ 
             1997, ApJ, 491, 153
\reference{} Kharchenko, N., Andruk, V., \& Schilbach, E.\ 1997,
             Astronomische Nachrichten, 318, 253 (KAS97)
\reference{} Lyng\r{a}, G.\ 1987, The Open Star Clusters Catalogue,
             5th edition
\reference{} Mathieu, R.~D., van den Berg, M., Torres, G., Latham, D.,
             Verbunt, F., \& Stassun, K.\ 2003, \aj, 125, 246
\reference{} Mochejska, B.~J., Stanek, K.~Z., Sasselov, D.~D.,
             \& Szentgyorgyi, A.~H.\ 2002, \aj, 123, 3460 (Paper~I)
\reference{} Monet, D.~G.~et al.\ 2003, \aj, 125, 984
\reference{} Patriarchi, P., Morbidelli, L., \& Perinotto, M.\ 2003,
             \aap, 410, 905
\reference{} Petersen, J.~O.~\& Christensen-Dalsgaard, J.\ 1999, \aap, 352,
             547 
\reference{} Piersimoni, A.~\&  et al.\ 1993, Memorie della Societa
             Astronomica Italiana, 64, 609 
\reference{} Pigulski, A., Kolaczkowski, Z., \& Kopacki, G.\ 2000,
             Acta Astronomica, 50, 113
\reference{} Rodr{\'{\i}}guez, E.\ 2002, ESA SP-485: Stellar Structure
             and Habitable Planet Finding, 319
\reference{} Rodr{\'{\i}}guez, E.~\& Breger, M.\ 2001, \aap, 366, 178 
\reference{} Schwarzenberg-Czerny, A.\ 1996, ApJ, 460, L107
\reference{} Stetson, P.~B.\ 1996, PASP, 108, 851
\reference{} Stetson, P.~B.\ 1987, PASP, 99, 191
\reference{} Szentgyorgyi, A.~H. et al.\ 2004, in preparation
\reference{} Templeton, M., Basu, S., \& Demarque, P.\ 2002, \apj,
             576, 963
\reference{} Warner, B.\ 1995, Cambridge Astrophysics Series,
             Cambridge, New York: Cambridge University Press
\end{references}
